\newif\ifprint
\newif\ifArxiv
\newif\ifAutomatica

\printtrue
\Arxivtrue

\ifArxiv
  \PassOptionsToClass{class/draft}{my-class/my-class}
\fi
\ifAutomatica
  \PassOptionsToClass{class/automatica}{my-class/my-class}
\fi

\documentclass{my-class/my-class}

\usepackage{my-goodcharter}

\usepackage{mathtools,amssymb}
\usepackage{wasysym}

\usepackage{bm}

\usepackage{my-semantic-logic-symbols}

\usepackage[
  shortcuts,
  allowbreakbefore,
]{extdash}

\makeatletter
\let\c@author\relax 
\makeatother
\usepackage[
  backend=biber,
  natbib=true,
  giveninits=true
]{biblatex}

\usepackage{titlecaps}
\Addlcwords{and as but for if nor or so yet}
\Addlcwords{a an the}
\Addlcwords{as at by for in of off on per to up via}

\DeclareFieldFormat{titlecase}{\titlecap{#1}}
\DeclareFieldFormat[online]{title}{\mkbibquote{#1\isdot}}
\usepackage[inline]{enumitem}

\usepackage{flafter} 
\usepackage{float} 

\usepackage{hyperref}
\usepackage[open,openlevel=2]{bookmark}

\usepackage{orcidlink}
\usepackage{uri}
\ifprint
\PassOptionsToPackage{mycolor=print}{my-cactus}
\else
\PassOptionsToPackage{mycolor=dark,mycoloroverride}{my-cactus}
\fi
\usepackage{my-cactus}

\usepackage{my-thm}

\ifAutomatica
  \usepackage{my-changes}
\fi

\DeclareMathOperator{\Obs}{O}
\DeclareMathOperator{\Con}{C}

\newcommand{\cD}{\mathcal{D}}
\DeclareMathOperator{\cdm}{cdm}
\newcommand{\dk}{\mathrm{dk}}
\newcommand{\cd}{\mathrm{cd}}

\mytitle{A Uniform Approach to Compare Architectures in Decentralized Discrete\-/Event Systems}
\mydate{\today}
\mydedicatory{
  This work was financially supported
  by the Natural Sciences and Engineering Research Council of Canada (NSERC)
  through a Collaborative Research and Development Grant
  together with General Dynamics Land Systems\--Canada
  and Defence Research and Development Canada (K.Ritsuka)
  and an NSERC Discovery Grant (K.Rudie).
}

\myauthor{K.~Ritsuka}
\myIEEEmembership{Student Member}
\myorcid{0000-0003-3713-5964}
\myemail{ritsuka314@queensu.ca}
\myaddressref{qu}

\myauthor{K.~Rudie}
\myIEEEmembership{Fellow}
\myorcid{0000-0002-8675-334X}
\myemail{karen.rudie@queensu.ca}
\myaddressref{qu}

\myaddress{qu}{
  The Department of Electrical and Computer Engineering and Ingenuity Labs Research Institute,
  Smith Engineering,\\
  Queen's University, Kingston, ON, Canada K7L 3N6
}

\mykeywords{
  Automata, Discrete event systems, Supervisory control, Agents and autonomous systems, Cooperative control
}

\myabstract{
  Solutions to decentralized discrete\-/event systems problems
  are characterized by the way local decisions are fused to yield a global decision.
  A fusion rule is colloquially called an \emph{architecture}.
  Current approaches do not
  provide a direct way
  to compare existing architectures.
  Determining whether an architecture is more permissive
  than another architecture had relied on producing examples \emph{ad hoc}
  and on individual inspiration that puts the conditions for solvability
  in each architecture into some form that admits comparison.
  In response to these research efforts,
  a method based on morphisms between graphs has been extracted to yield a uniform approach
  to compare the permissiveness of the architectures.
}

\begin{document}

\mymaketitle

\section{Introduction}

Studies of decentralized supervisory control problems
generally include topics of control architecture design,
supervisor existence, supervisor realization, and complexity analysis.

For architecture design,
existing works include \citep{Cieslak1988,Rudie1992,Prosser1997,Yoo2002,Yoo2004,Kumar2007,Chakib2011}.
Since one of the goals in designing new architectures
is to develop an architecture that is more permissive than existing ones,
it is necessary to compare architectures by
examining the class of closed-loop behaviour synthesisable under that architecture.

The current strategy for comparing the permissiveness of two architectures
is as follows.
First, a characterization for the synthesisable behaviours
are given for each of the two architectures.
The characterizations are used in the following way:
\begin{enumerate}
  \item To show that one architecture is at least as permissive as the other,
  one shows that the latter's characterization logically implies that of the former.
  \item To show that an architecture synthesizes some behaviour not synthesizable by the other,
  one constructs an example and verifies/disproves the respective characterizations.
  In principle, one can show logical non-implication without constructing an explicit example,
  but it is actually harder to achieve.
  \item To show that an architecture is \emph{strictly} more permissive,
  one has to perform both of the two tasks above.
  \item To show that two architectures are incomparable,
  one has to perform the second task twice.
\end{enumerate}

The drawbacks of the traditional approach are present in all of the tasks:
deriving characterizations (and their decision procedures),
demonstrating logical implications,
finding counterexamples,
and demonstrating a counterexample satisfies/doesn't satisfy the characterizations.
All of these tasks rely heavily on case-by-case ingenuity;
moreover the sheer number of tasks also makes it non-trivial to verify
that a candidate architecture is, indeed, more permissive than an existing one.

A consequence of the challenges of the traditional approach can be seen
in establishing the relation between the disjunctive architecture \citep{Prosser1997}
and the conjunctive architecture \citep{Cieslak1988,Rudie1990}.
It took years until \citet[Proposition 3]{Yoo2002} showed
that the two architectures are incomparable,
using the traditional paradigm.

The framework presented by \citet{RitsukaEpistemic} can be used to
address some of the issues above.
It provides a more methodologically\-/guided approach to derive characterizations,
and in many cases these characterizations are easily comparable and decidable.
However, there has been no other work making other parts of the traditional approach easier,
nor one that provides an entirely different approach.

This paper presents a uniform approach
for easy and direct comparison of decentralized architectures.
We proceed to it as follows:
\begin{itemize}
  \item \Cref{sec:preliminary} recalls a reduction of control problems
  to observation problems,
  which is used to simplify the discussion without loss of generality.
  \item \Cref{sec:o2d} casts the characterizations into a uniform and abstract presentation,
  which is only used to establish the correctness of our main result.
  \item Finally, \cref{sec:d2d} presents the novel approach for comparing permissiveness of architectures,
  without the need to pass indirectly through characterizations and counterexamples.
\end{itemize}

\section{Decentralized Problems}
\label{sec:preliminary}

This section first formally defines decentralized control problems.
Then we make use of a fact by \citet{RitsukaCorrespondence}
to simplify control problems to observation problems
to make the discussion terser.

\begin{prob}[Decentralized Control Problem]
  Given an alphabet $\Sigma$,
  controllable alphabets $\Sigma_{c,i} \subseteq \Sigma$,
  observation functions $P_i$ over $\Sigma^*$
  for agents $i \in \nset = \set{1,\allowbreak \dots,\allowbreak n}$,
  and languages $K\subseteq L \subseteq \Sigma^*$.
  We use the notations
  $\Sigma_c = \Union_{i \in \nset} \Sigma_{c,i}$ and
  $\Sigma_u = \Sigma - \Sigma_c$.
  The Control Problem is to construct
  \emph{local decision functions} $f^\sigma_i$ over $P[\Sigma^*]$
  (also informally called \emph{supervisors}/\emph{agents}),
  one for each $i \in \nset$ and $\sigma \in \Sigma_{c,i}$,
  and \emph{fusion rules} $f^\sigma$,
  one for each $\sigma \in \Sigma_c$,
  such that
  \begin{align*}
    & \All{s \in K}
    \\
    & \quad
      \begin{alignedat}[t]{2}
        & s\sigma \in K     && \implies f^\sigma(\setidx{f^\sigma_iP_i(s)}{\setcomp{i}{\sigma \in \Sigma_{c,i}}}) = 1
        \\ {}\land{}
        & s\sigma \in L - K && \implies f^\sigma(\setidx{f^\sigma_iP_i(s)}{\setcomp{i}{\sigma \in \Sigma_{c,i}}}) = 0.
      \end{alignedat}
  \end{align*}
\end{prob}

To avoid trivially unsolvable instances,
always assume $K\Sigma_u \sect L \subseteq K$.
This condition is called \emph{controllability}.

An instance of control problem is denoted by
$\Con(L,\allowbreak K,\allowbreak \setidx{P_i}{i \in \nset},\allowbreak \setidx{\Sigma_{c,i}}{i \in \nset})$,
or more succinctly,
$\Con(L,\allowbreak K,\allowbreak P_i,\allowbreak \Sigma_{c,i})$.

If the same fusion rule $f$ is determined for all $\sigma \in \Sigma_c$,
then the problem is denoted by $\Con(f, L,\allowbreak K,\allowbreak P_i,\allowbreak \Sigma_i)$.

\Citet{RitsukaCorrespondence} observed that a control problem
can be separated into independent sub-problems,
one for each $\sigma \in \Sigma_c$,
where each sub-problem can be recast
into a simpler form as an observation problem.

\begin{prob}[Decentralized Observation Problem]
\label{prob:dop}%
  Given an alphabet $\Sigma$,
  observation functions $P_i$ over $\Sigma^*$
  for agents $i \in \nset = \set{1, \dots, n}$,
  and languages $K\subseteq L \subseteq \Sigma^*$.
  The observation problem is to construct
  \emph{local decision functions} $f_i$
  (also informally called \emph{observers}/\emph{agents}),
  one for each $i \in \nset$,
  and a \emph{fusion rule} $f$,
  such that
  \begin{align*}
    & \All{s \in L} 
    \\
    & \quad
      \begin{alignedat}[t]{2}
        & s \in K     && \implies f(f_1P_1(s), \dots, f_nP_n(s)) = 1
        \\ {}\land{}
        & s \in L - K && \implies f(f_1P_1(s), \dots, f_nP_n(s)) = 0.
      \end{alignedat}
  \end{align*}
\end{prob}

An instance of the observation problem is denoted by
$\Obs(L, K, \setidx{P_i}{i \in \nset})$
or more succinctly,
$\Obs(L, K, P_i)$.

If the fusion rule $f$ is determined,
then the problem is denoted by $\Obs(f, L, K, P_i)$.

\begin{thm}[\citep{RitsukaCorrespondence}]
  Define the following two languages
  \begin{equation}
    \label{eq:LK}
    \begin{aligned}
      L_\sigma &= \setcomp{s \in K}{s\sigma \in L} \\
      K_\sigma &= \setcomp{s \in K}{s\sigma \in K},
    \end{aligned}
  \end{equation}
  then
  solving a control problem $\Con(f, L, K, P_i, \Sigma_{c,i})$
  is equivalent to 
  solving the observation problems
  $$\setidx{\Obs(f, L_\sigma, K_\sigma, \setidx{P_i}{i \in \nset_\sigma})}{\sigma \in \Sigma_c}.$$
\end{thm}

Note that in the theorem above,
the languages $L_\sigma$ and $K_\sigma$
need not be prefix\-/closed even if $L$ and $K$ are.

Only for the purpose of being illustrative,
we will use natural projections to $\Sigma_{o,i}^*$ as our $P_i$ in the examples,
where $\Sigma_{o,i} \subseteq \Sigma$ are the observed alphabets.
Nonetheless, our main result is not subject to this restriction.

The problem of comparing architectures can then be stated as follows.

\begin{prob}[Permissiveness]
  Give a fusion rule $f$
  let $\mathcal{O}_f$
  denote the set of Observation Problems $\Obs(f, L, K, P_i)$
  that have a solution.
  This set measures the permissiveness of the fusion rule $f$.

  Then, given two fusion rules $f$ and $g$,
  the problem of comparing their permissiveness
  is determining the relationship between
  the two sets $\mathcal{O}_f$ and $\mathcal{O}_g$:
  if all problems solvable by $f$ are solvable by $g$,
  i.e., if $\mathcal{O}_f \subseteq \mathcal{O}_g$,
  then $g$ is at least as permissive as $f$.
  If the inclusion is strict,
  then $g$ is strictly more permissive than $f$.
  If there are problems solvable by $f$ but not $g$
  and vice versa,
  i.e., if $\mathcal{O}_f$ and $\mathcal{O}_g$ are incomparable
  by set inclusion,
  then we say that $f$ and $g$ are incomparable.
\end{prob}

Before presenting our main result,
we will need the following definitions to handle functions on $n$-tuples.

\begin{defn}
  For $n$ functions $g_1, \dots, g_n\colon A \to B$,
  define the \emph{broadcasting application} (as broadcasting $x \in A$ to each $g_i$)
  $$\langle g_1, \dots, g_n \rangle\colon A \to B^n$$
  where
  $$x \mapsto (g_1(x), \dots, g_n(x)),$$
  and the \emph{pair-wise application} (of $x_i$ and $g_i$)
  $$(g_1, \dots, g_n)        \colon A^n \to B^n$$
  where
  $$(x_1, \dots, x_n) \mapsto (g_1(x_1), \dots, g_n(x_n)).$$
  For compactness,
  we may write $\langle g_1, \dots, g_n \rangle (x)$ as $\langle g_i \rangle x$
  and $(g_1, \dots, g_n)(x)$ as $(g_i) x$.
\end{defn}

We also need the following definitions regarding
equivalence relations induced by functions,
partitions of sets,
and refinement of equivalence relations.

\begin{defn}
  A function $f$ induces an equivalence relation as
  $\ker f = \setcomp{(x, y)}{f(x) = f(y)}$.

  Finally,
  say a relation $R$ refines a relation $T$,
  written as $R \leq T$,
  if $(x,y) \in R$ whenever $(x,y) \in T$.
\end{defn}

For two strings $s_1$ and $s_2$ in $L$
such that $P_i(s_1) = P_i(s_2)$,
necessarily $f_iP_i(s_1) = f_iP_i(s_2)$ for all $i \in \nset$.
We call this fact \emph{feasibility}.
Feasibility can be described in terms of
refinement of function kernels:
$\ker (P_1, \dots, P_n) \leq \ker (f_1P_1, \dots, f_nP_n)$,
i.e.,
the first kernel refines the second.

\section{A Uniform Approach to Derive Problem Solvability Characterization from a Given Fusion Rule}
\label{sec:o2d}

Our aim is to derive a uniform approach to
compare decentralized architectures directly.
We do so by first giving a uniform way
to derive problem solvability characterization,
from which we will then derive our approach
for comparing architectures.
In our approach, we
describe a decentralized problem
as an \emph{observation graph}
and a solution based on a fusion rule $f$
as a \emph{decision graph}.
The problem can be expressed as
finding a way of folding the observation graph
into the decision graph,
which will be formally expressed in terms of graph morphism.
Then the problem solvability condition can be thought
as determining if the decision graph has the capacity
to embed the observation graph.

\begin{defn}[Observation Graph]
  Define the symmetric relations $\sim_{N}$ on $L$,
  so that $s \sim_{N} t$
  if and only if the two tuples
  $\langle P_i \rangle s = (P_1(s), \dots, P_n(s))$ and
  $\langle P_i \rangle t = (P_1(t), \dots, P_n(t))$
  differ by exactly the components indexed by ${N}$.
  Formally,
  \[
    {\sim_N} = \setcomp{(s, t) \in L \times L}{P_i(s) \ne P_i(t) \iff i \in N}.
  \]
  The relations $\sim_{N}$ reflect that
  exactly the agents in $N$
  observe $s$ and $t$ differently.
  We may consider $L$ and $\sim_{N}$ to form an undirected graph
  $(L, \sim)$,
  which we will call an \emph{observation graph}.
  We denote the observation graph also with $L$.
  We consider the graph as a complete graph
  where edges are coloured by subsets of $\nset$.
  We also colour a node $s$ by the truth value of $s \in K$.
\end{defn}

The equivalence relation $\sim_\emptyset$
is exactly the kernel
$\ker \langle P_i \rangle$.

\begin{ex}
  \label{ex:problem}
  Consider the observation problem with two agents
  where
  $L = \{a, b, ab, bb\}$,
  $K = b$,
  $\Sigma_{o,1} = \{a\}$ and
  $\Sigma_{o,2} = \{b\}$.
  This example is derived from \citep[Fig. 1]{Rudie1992}.
  We depict the observation graph
  as in \cref{fig:example}.
  Each node is labelled by strings $s \in L$, $P_1(s)$, and $P_2(s)$,
  vertically stacked in that order.
  Vertical/blue/dotted lines denote relation $\sim_1$;
  horizontal/red/dashed lines denote relation $\sim_2$; and
  diagonal/purple/solid lines denote relation $\sim_{1,2}$.
  The relation $\sim_\emptyset$ happens to be the identity relation
  for this example
  and is omitted from the graph.
  Red/singly-bordered nodes indicate strings in $L - K$, and
  green/doubly-bordered nodes indicate string in $K$.
  \begin{figure}
    \begin{center}
      \color{black}
      \includegraphics{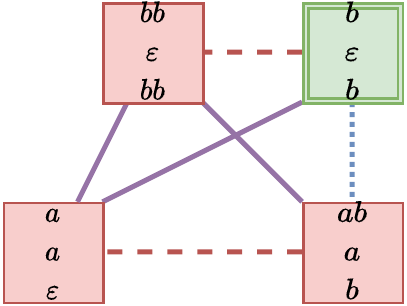}
    \end{center}
    \caption{Observation graph for the observation problem in \cref{ex:problem}.}
    \label{fig:example}
  \end{figure}
\end{ex}

The observation graphs are essentially a more compact alternative
to the Kripke structures used in the works employing epistemic logic interpretations
for decentralized problems \citep{Ricker2000,Ricker2007,RitsukaEpistemic}.
The relations $\sim_N$ capture various notions of group knowledge,
e.g., what is expressed
by distributed knowledge and
by the ``everybody knows'' operators in epistemic logic
\citep{Fagin2004}.

Without loss of generality,
suppose that the local decision functions $f_i$
all have codomain $D$,
for otherwise we can simply take
$D = \Union_{i \in \nset} \cdm(f_i)$,
where $\cdm(f_i)$ stands for the codomain of $f_i$.
Hence, the domain of $f$
is a subset of $D \times \dots \times D$ ($n$ times).

Suppose that $f$ is only defined over a certain collection $\cD$
of combinations of local decisions%
$(d_1, \dots, d_n) \in D \times \dots \times D$,
i.e., $\cD = \operatorname{dom}(f) \subseteq D \times \dots \times D$.
The size of $\cD$ roughly reflects the capacity of $f$,
so that if $|\cD| = 1$, $f$ is a constant $1$ or $0$,
and if $\cD$ is large enough for a problem at hand,
$f$ is virtually unconstrained.

The size of $\cD$ alone does not fully capture
the capacity of the fusion rule.
What we need additionally
is the following.

\begin{defn}[Decision Graph]
  For each subset $N \subseteq \nset$,
  define symmetric relations $\approx_{N}$ on $\cD$,
  so that
  $(d_1, \dots, d_n) \approx_{N} (d'_1, \dots, d'_n)$
  when the two tuples differ by exactly the components indexed by $N$.
  Formally,
  \ifArxiv
  \[
    {\approx_{N}} =
    \setcomp{((d_1, \dots, d_n), (d'_1, \dots, d'_n)) \in \cD
    \times \cD}{d_i \ne d'_i \iff i \in N}
  \]
  \fi
  \ifAutomatica
  \begin{align*}
    {\approx_{N}} =
    \setcomp{&((d_1, \dots, d_n), (d'_1, \dots, d'_n)) \in \cD
    \times \cD\\}{{}&d_i \ne d'_i \iff i \in N}
  \end{align*}
  \fi
  The relations $\approx_{N}$ reflect that
  exactly the agents in $N$
  have their decisions $d_i'$ differ from $d_i$.
  We may consider $\cD$ and $\approx_{N}$ to form an undirected graph
  $(\cD, \approx)$,
  which we will call a \emph{decision graph}.
  We denote the decision graph also with $\cD$.
  We consider the graph as a complete graph
  where edges are coloured by subsets of $\nset$.
  We also colour nodes by the values of $f$ at the nodes (0 or 1).
\end{defn}

\begin{ex}
  \label{ex:conj}

  The traditional way of describing the conjunctive architecture
  is by taking $D = \set{1, 0}$ and $f = \land$ is the Boolean conjunction \citep{Rudie1992}.
  For a problem with two agents,
  the decision graph can be depicted
  as in \cref{fig:conj}.
  Similar to how we depicted the observation graph in \cref{ex:problem},
  vertical/blue/dotted lines denote relation $\approx_1$;
  horizontal/red/dashed lines denote relation $\approx_2$; and
  diagonal/purple/solid lines denote relation $\approx_{1,2}$.
  The relation $\approx_\emptyset$ is the identity relation
  and is omitted from the graph.
  Red/singly-bordered nodes indicate fused decision being $0$ and
  green/doubly-bordered nodes indicate fused decision being $1$.
  \begin{figure}
    \begin{center}
      \color{black}
      \includegraphics{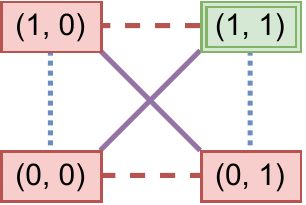}
    \end{center}
    \caption{Decision graph for the conjunctive architecture.}
    \label{fig:conj}
  \end{figure}
\end{ex}

Solving an observation problem is essentially
finding a way to fold the observation graph into the decision graph.
We proceed to describe such foldings formally
as a type of graph morphisms.

The local decision functions $f_i$
can be seen as a mapping from
$P_i(L)$ to $\cD$,
subject to the following requirement.
For any strings $s$ and $s'$ in $L$,
let $N$ be the (unique) set such that
$$s \sim_{N} s'.$$
If
\[\begin{array}{@{}r@{}l@{}l@{}l@{}l@{}r@{}r@{}c@{}l@{}c@{}l@{}}
  (&P_1(s )&, \dots, &P_n(s )&) &{}\xmapsto{(f_i)}{} &(&f_1P_1(s )&, \dots, &f_nP_n(s )&) \\
   &       &         &       &  &              {}={} &(&d _1      &, \dots, &d _n      &) \\
  (&P_1(s')&, \dots, &P_n(s')&) &{}\xmapsto{(f_i)}{} &(&f_1P_1(s')&, \dots, &f_nP_n(s')&) \\
   &       &         &       &  &              {}={} &(&d'_1      &, \dots, &d'_n      &),
\end{array}\]
then,
with letting $N'$ be the set such that
$(d_1, \dots, d_n) \approx_{N'} (d'_1, \dots, d'_n)$,
the mapping $g = \langle f_iP_i \rangle = s \mapsto \tup{f_1P_1(s), \dots, f_nP_n(s)}$ must satisfy
the following conditions \hypertarget{GM}{GM}:
\begin{enumerate}[leftmargin=*,label=GM\-/\arabic*:,ref=GM\-/\arabic*]
  \item {Node-Colour Preserving}
    \label{GM-1}

    The mapping $g$ preserves node colouring,
    that is, $g$ achieves the desired fused decision.

  \item {Edge-Colour Intensive}
    \label{GM-2}

    $N \supseteq N'$,
    i.e., only agents
    observing $s$ and $t$ differently
    can issue different decisions for the two strings,
    though they do not necessarily have to.
    In other words,
    the mapping $g$ may drop some edge colours,
    but may not add any.
    This property captures feasibility.
\end{enumerate}
We call a mapping satisfying \hyperlink{GM}{GM}
a (graph) morphism for the purpose of this work.

We claim that converse also holds:
a morphism satisfying the two conditions above
gives a solution to the problem.

We capture the foregoing in the following theorem.

\begin{thm}
  \label{thm:o2d}
  An Observation Problem $\Obs(f, L, K, P_i)$
  has a solution
  if and only if
  there exists a morphism 
  from the observation graph to the decision graph (representing the architecture's fusion rule)
  satisfying the morphism conditions \hyperlink{GM}{GM}.
\end{thm}

\begin{pf}
  ($\implies$):
  By the discussion preceding the theorem,
  $g$ satisfies \hyperlink{GM}{GM},
  as that is how the definition of \hyperlink{GM}{GM} was motivated.
  
  ($\impliedby$):
  Suppose that there is a morphism $g$ satisfying the morphism conditions \hyperlink{GM}{GM}.
  Then a solution can be constructed as follows.
  For each string $s \in L$,
  let $(d_1, \dots, d_n) = g(s)$,
  and let $f_iP_i(s) = d_i$ for $i \in \nset$.
  Since $g$ is edge-colour intensive (\labelcref{GM-2}),
  the functions $f_i$ are well-defined,
  i.e.,
  if there were $s$ and $s'$ such that $P_i(s) = P_i(s')$,
  then $f_iP_i(s) = f_iP_i(s') = d_i$.
  Since $g$ preserves node colours (\labelcref{GM-1}),
  $f_i$ solves the problem.
  In other words,
  there must exist $f_i$ such that
  $g = \langle f_iP_i \rangle$.
\end{pf}

\begin{ex}
  The problem in \cref{ex:problem} is solvable in the conjunctive architecture (\cref{ex:conj}),
  as we can construct the morphism
  depicted in \cref{fig:solution}.
  \begin{figure}
    \begin{center}
      \color{black}
      \includegraphics{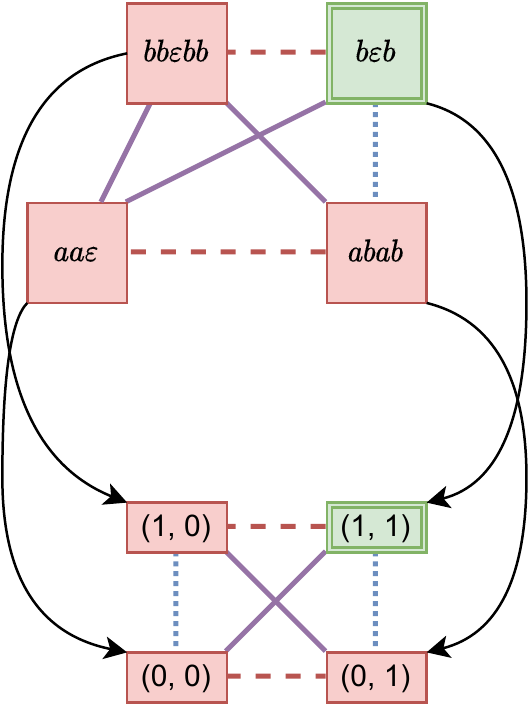}
    \end{center}
    \caption{Graph morphism from the observation graph in \cref{fig:example} to the decision graph in \cref{fig:conj}.}
    \label{fig:solution}
  \end{figure}
  Notice how 
  the leftmost diagonal/purple/solid edge in the graph on top
  loses its redness/horizontalness
  and become a vertical/blue/dotted edge in the graph on the bottom.
  All other node/edge colours do not change through the morphism.
  
  The morphism gives a solution to the problem.
\end{ex}

\section{A Uniform Approach to Compare Fusion Rules}
\label{sec:d2d}

The traditional way to compare two fusion rules
is by first obtaining a characterization of
problem solvability with each of the fusion rules,
and then determining whether one characterization logically entails the other.
In light of the discussion in the previous section,
we can obtain a more direct way to compare fusion rules
without deriving characterizations of problem solvability first.

From the respective definitions
from the previous section,
one can see that
there is no formal distinction between
observation graphs and decision graphs.
Consequently,
we have the following result.

\begin{thm}
  \label{thm:d2d}
  Given fusion rules $f$, $f'$
  and their respective decision graphs
  $\cD$, $\cD'$,
  the fusion rule $f'$ is more permissive
  than $f$
  (i.e., if $\Obs(f, L, K, P_i) \in \mathcal{O}_f$,
  then $\Obs(f', L, K, P_i) \in \mathcal{O}_f'$)
  if and only if
  there is a graph morphism
  from $\cD$ to $\cD'$
  satisfying the graph morphism conditions \hyperlink{GM}{GM}.
\end{thm}

The statement of \cref{thm:d2d} makes it necessary to view
the decision graph $(\cD, \approx_N)$ as an observation graph.
Hence we need the following lemma.

\begin{lem}
  \label{lem:d2o}
  For every decision graph $(\cD, \approx_N)$,
  there is an isomorphic observation graph,
  where the isomorphism preserves node- and edge-colouring.
\end{lem}

To avoid distraction,
we postpone the proof of \cref{lem:d2o}
until \cref{sec:App}.

Now we are ready to prove \cref{thm:d2d}.

\begin{pf}
  ($\impliedby$):
  Suppose there is a morphism $g \colon \cD \to \cD'$
  satisfying the morphism conditions \hyperlink{GM}{GM}.
  Consider an arbitrary observation problem solvable with the fusion rule $f$.
  By \cref{thm:o2d},
  there is a morphism $h \colon L \to \cD$
  satisfying the morphism conditions \hyperlink{GM}{GM}.
  Then $g \circ h \colon L \to \cD'$ is a morphism
  that satisfies the morphism conditions \hyperlink{GM}{GM},
  and therefore,
  by \cref{thm:o2d},
  solves the observation problem.
  Thus,
  all problems solvable with $f$
  are also solvable with $f'$.
  
  ($\implies$):
  Suppose that all observation problems solvable with the fusion rule $f$
  are solvable with $f'$.
  View $\cD$ as an observation problem according to \cref{lem:d2o}.
  The observation problem must be solvable with $f$ by the isomorphism,
  hence by assumption it must be solvable with $f'$.
  By \cref{thm:o2d}, there must be a morphism $h$
  from the observation graph to $\cD'$.
  Because the observation graph is isomorphic to $\cD$,
  and the composition of an isomorphism
  and a morphism \--- both satisfying \hyperlink{GM}{GM} \---
  satisfies \hyperlink{GM}{GM},
  we obtain the desired morphism from $\cD$ to $\cD'$.
\end{pf}

We now illustrate how two architectures
can be directly compared with our approach.
We first show how one architecture can be determined
to be strictly more general than another.

\begin{ex}
  \label{ex:cpda}
  Consider the architecture
  in which the local decisions available are $\set{0, 1, \dk}$,
  where $\dk$ stands for ``don't know''.
  The associated fusion rule outputs $0$ whenever a $0$ local decision is present,
  and $1$ whenever a $1$ local decision is present,
  and is undefined when
  either there are conflicting local decisions (both $0$ and $1$ are present),
  or all supervisors don't know (all supervisors are confused).
  
  We called this architecture the C\&P$\wedge$D\&A architecture
  in our earlier work \citet{RitsukaEpistemic}.
  The term comes from the relationship of this architecture
  with the C\&P (conjunctive and permissive) architecture \citep{Rudie1992}
  and the D\&A architecture (disjunctive and anti\-/permissive) \citep{Prosser1997}.
  
  The decision graph is depicted in \cref{fig:cpda},
  where grey/dash-bordered nodes indicate disallowed combination of local decisions
  and hence are not formally in the decision graph $\cD$.
  \begin{figure}
    \begin{center}
      \color{black}
      \includegraphics{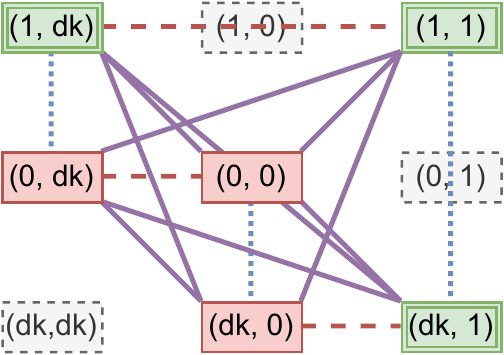}
    \end{center}
    \caption{Decision graph for the C\&P$\wedge$D\&A architecture.}
    \label{fig:cpda}
  \end{figure}
  The C\&P$\wedge$D\&A architecture is known to be weaker than the
  C\&P architecture (which we have been calling the ``conjunctive architecture'').
  This fact can be readily demonstrated by giving a decision graph morphism,
  which sends all nodes in the dashed box to $(1,1)$.
  \Cref{fig:cpda-cp} depicts such a morphism,
  where for representation purpose we no longer make use
  of horizontalness/verticalness to denote edge colouring.
  \begin{figure}
    \begin{center}
      \color{black}
      \includegraphics{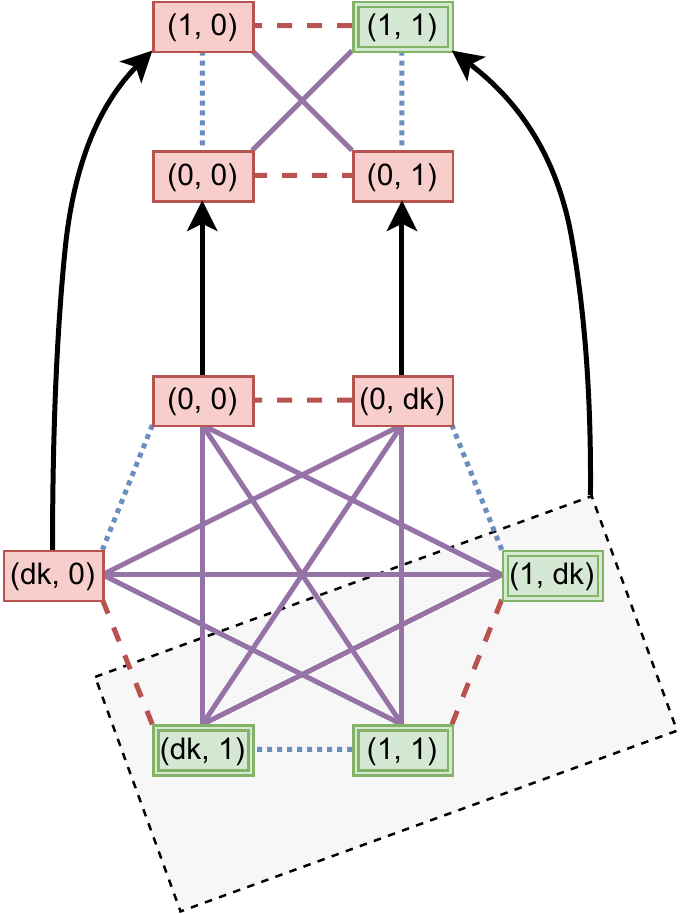}
    \end{center}
    \caption{
      Graph morphism from
      the decision graph for the C\&P$\wedge$D\&A architecture (bottom)
      to the decision graph for the C\&P architecture (top).}
    \label{fig:cpda-cp}
  \end{figure}

  One can also see that there can be no morphism going in the other direction,
  as there is no green/doubly\-/bordered node in the bottom graph
  having both a red/dashed edge and a blue/dotted edge
  to a red/singly\-/bordered node,
  which is necessary for the node $(1,1)$ in the top graph.
\end{ex}

The following example shows
how two seemingly different architectures
can be determined to be equivalent.

\begin{ex}
  An alternative way to describe the conjunctive architecture
  is by using three decisions $\set{0, 1, \cd}$,
  where $\cd$ is to be interpreted as a conditional decision,
  so that the fusion rule outputs $1$ when only the conditional decision is present,
  and otherwise behaves identically to the fusion rule
  in the C\&P$\wedge$D\&A architecture as given in \cref{ex:cpda}
  (although we renamed the decision $\dk$ to $\cd$).
  The decision graph is depicted in \cref{fig:cp2} without edges for compactness.
  \begin{figure}
    \begin{center}
      \color{black}
      \includegraphics{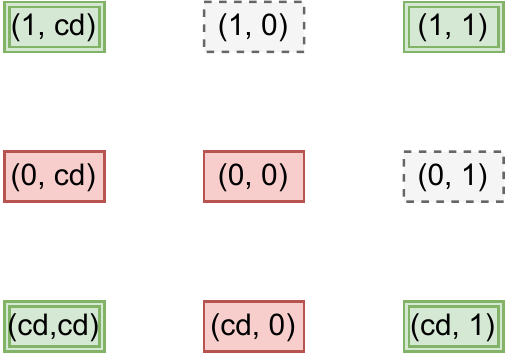}
    \end{center}
    \caption{Alternative decision graph for the conjunctive architecture.}
    \label{fig:cp2}
  \end{figure}

  It is easy to check that this architecture is indeed equivalent
  to the conjunctive architecture.
  Since the graph would be too complex to draw,
  we describe the morphisms verbally.
  The morphism $h$ to the conjunctive architecture
  is like the morphism from the C\&P$\wedge$D\&A architecture to the conjunctive architecture
  as given in the previous example,
  where all green/doubly\-/bordered nodes are sent to $(1, 1)$.
  Unlike the C\&P$\wedge$D\&A architecture,
  now we have a morphism $g$ from the conjunctive architecture:
  red/singly\-/bordered nodes are mapped by reversing $h$,
  where the only green/doubly\-/bordered node $(1, 1)$ is mapped to $(\cd, \cd)$.
  That is, in the C\&P architecture,
  the decision $1$ can be interpreted as a conditional decision,
  which aligns with the interpretation in \citep{RitsukaEpistemic}.

  The foregoing shows that
  there may exist two architectures
  that are equivalent in the sense of having morphisms
  in both directions,
  for example, the two architectures depicted in \cref{fig:conj,fig:cp2}.
  However,
  although the architecture in \cref{fig:cp2}
  has more nodes than that of \cref{fig:conj},
  the redundancy in this case serves a purpose:
  the original formulation of the conjunctive architecture by \citet{Rudie1992}
  essentially forced the decision $1$ (enable)
  to stand for both an agent actively enabling an event
  because the agent knew the event was legal,
  and passively enabling the event when the agent didn't know if the event was legal.
  To understand the meaning behind agents' behaviours,
  it is useful to separate out the roles played by a decision,
  which is exactly what the architecture in \cref{fig:cp2} does.
\end{ex}

The following example shows
how two architectures can be determined to be incomparable.

\begin{ex}
  \label{ex:cp-da}
  Recall the decision graph for the conjunctive architecture
  (where $f = \land$)
  in the left part of \cref{fig:conj-disj}.
  Compare it with the disjunctive architecture \citep{Prosser1997},
  also known as the D\&A architecture 
  (where $f = \lor$),
  whose decision graph is depicted in the right part of \cref{fig:conj-disj}.
  \begin{figure}
    \begin{center}
      \color{black}
      \includegraphics{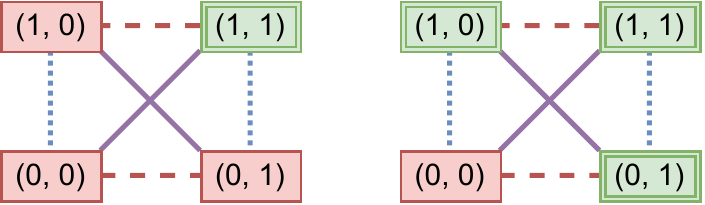}
    \end{center}
    \caption{
      Decision graph for the conjunctive architecture recalled in the left,
      with the decision graph for the disjunctive architecture in the right.
    }
    \label{fig:conj-disj}
  \end{figure}
  
  One can see that there can be no morphism from left to right,
  as there is no green/doubly\-/bordered node in the right graph
  having both red/dashed and blue/dotted edges to red/singly\-/bordered nodes,
  which is necessary for the node $(1,1)$ in the left graph.
  By a similar argument over the node $(0, 0)$ in the right graph,
  one can see that there can be no morphism from right to left either.
  This is sufficient to determine that the conjunctive architecture
  and the disjunctive architecture are incomparable.
\end{ex}

The fact that the conjunctive and disjunctive architectures are incomparable
was not known at the time of the original formulation
of the disjunctive architecture \citep{Prosser1997},
but was only discovered five years after by \citet{Yoo2002}.
They had to explicitly produce elements in 
$\mathcal{O}_{\land} \setminus \mathcal{O}_{\lor }$ and
$\mathcal{O}_{\lor } \setminus \mathcal{O}_{\land}$,
which involves finding two elements and verifying/disproving four set\-/memberships.
In contrast, by applying \cref{thm:d2d},
the incomparability is apparent as illustrated in \cref{ex:cpda}.

\section{Conclusion}

We proposed two tools
for studying decentralized observation problems:
observation graphs and decision graphs.
The decision graphs alone provide a systematic approach
to directly compare decentralized architectures.
Together with observation graphs,
we have systematic approaches
to derive problem solvabilities and solutions.

As we can see in the development of decentralized observation problems,
the earlier works propose verifiable characterizations for problem solvability
and computable algorithms to construct solutions
\citep{Cieslak1988,Rudie1992,Prosser1997,Yoo2002,Yoo2004},
but subsequent works
can no longer provide computable solvability characterizations,
let alone algorithms to construct solutions
\citep{Kumar2007,Chakib2011}.
Said differently,
finding a graph morphism may be hard,
but verifying a witness could be easier.
Specifically,
when the graphs involved are finite,
the problem can be solved in nondeterministic polynomial time,
but has proofs verifiable in polynomial time.
When the graphs are infinite,
the problem can be undecidable,
but proofs can be verified.
This suggests
that in a situation where the solvability characterization becomes undecidable,
one should attempt to prove the characterization instead.
Moreover,
when a solution is ``finite'' in some sense,
e.g., the solution is described by finite state automata,
it remains verifiable.

In summary,
fusion rules with unbounded numbers of decisions present challenges
for finding graph morphisms.
Nonetheless,
for fusion rules with finite, bounded numbers of decisions,
our work provides a direct and easy approach to compare
the corresponding architectures.

\appendix
\section{Appendix}
\label{sec:App}

Here we provide a proof of \cref{lem:d2o}.

\begin{pf}
  We construct an observation problem whose observation graph
  is isomorphic to $\cD$.
  Recall that $\cD \subseteq D \times \cdots \times D$.
  Construct the following problem.
  Take $\Sigma = \setcomp{(d, i)}{d \in D \land i \in \nset}$
  as our alphabet,
  where each symbol consists of a decision $d$,
  tagged by an agent $i$,
  where $(d, i)$ is alternatively written as $d^i$.
  Associate to each node $v = (d_1, \cdots, d_n)$ in $\cD$
  the string $s_v = d_1^1 \cdot\cdots\cdot d_n^n$.
  This association is bijective.

  To preserve node-colouring,
  let $s_v \in K$ if $v$ is coloured green/doubly\-/bordered,
  and $s_v \in L - K$ if $v$ is coloured red/singly\-/bordered.
  To preserve edge-colouring,
  take $\Sigma_{o,i} = \setcomp{d^i}{d \in D}$,
  so that $P_i(s_v) = d_i^i$.

  Note that, 
  when the set of available decisions $D$ is countably infinite,
  the alphabet is countably infinite.
  The need for an infinite alphabet can be eliminated,
  as we can encode symbols in an infinite alphabet
  in terms of a finite alphabet.
  Using a finite alphabet instead
  however requires more sophistication
  in specifying the desired observability.
  Since we have assumed that the decision set is enumerable,
  let function $\llcorner \cdot \lrcorner: D \to \mathbb{N}$
  be the enumeration of decisions in natural numbers.
  This enumeration function allows us to speak of the ``$j$-th'' decision in the set $D$.
  First take $\Sigma = \Union_{i \in \nset}\set{0_i, 1_i}$
  and $\Sigma_{o,i} = \set{0_i, 1_i}$.
  Then associate to each node $v = (d_1, \cdots, d_n)$ in $\cD$
  the string $s_v = 0_1^{\llcorner d_1 \lrcorner}1_1 \cdots 0_n^{\llcorner d_n \lrcorner}1_n$,
  so that $P_i(s_v) = 0_i^{\llcorner d_i \lrcorner}1_i$.
  The intention of the encoding is that
  $0$ enumerates decisions in unary notation,
  $1$ marks the endings of code words,
  and subscripts impose observabilities.
  In other words,
  $0_i^{\llcorner d_i \lrcorner}$ means
  a string of $0$'s of length $\llcorner d_i \lrcorner$.
  The idea is that if $d_i$ is the $j$'th decision in the set $D$,
  then it gets encoded by $j$ $0$'s followed by $1$.
  
  Since the enumeration $\llcorner \cdot \lrcorner$ is injective,
  the encoding $d_i \mapsto 0_i^{\llcorner d_i \lrcorner}1_i$
  is also injective.
  Moreover, the encoding is prefix-free
  and hence instantaneously and uniquely decodable,
  i.e., the association to $v$ of $s_v$ is one-to-one.
\end{pf}

We illustrate the methodology in the proof of \cref{lem:d2o}
on the following example.
The example uses a finite decision set,
so that we can display the observation graph of our example
however, the methodology is the same for infinite but countable sets $D$.

\begin{ex}
  \label{ex:conj-d-to-o}
  The decision graph of the conjunctive architecture in \cref{ex:conj}
  can be seen as the observation graph of the following problem.
  Let the enumeration function $\llcorner \cdot \lrcorner$
  send the symbol $0$ to the number $0$
  and the symbol $1$ to the number $1$.
  With $\nset = \set{1, 2}$,
  let $\Sigma = \set{0_1, 1_1, 0_2, 1_2}$,
  $\Sigma_{o,1} = \set{0_1, 1_1}$, and
  $\Sigma_{o,2} = \set{0_2, 1_2}$.
  Let $L = \set{1_1 1_2, 0_11_1 1_2,1_1 0_21_2, 0_11_1 0_21_2}$ and $K = \set{0_11_1 0_21_2}$.
  Then the observation graph is
  depicted in \cref{fig:d2o}.
  \begin{figure}
    \begin{center}
      \color{black}
      \includegraphics{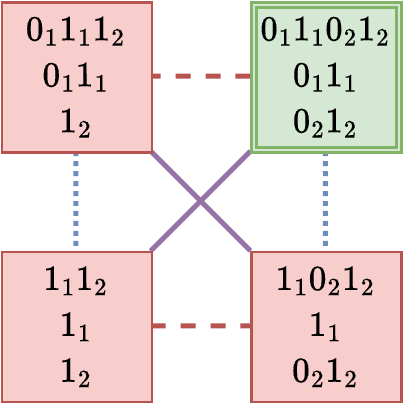}
    \end{center}
    \caption{An observation graph that is isomorphic to the decision graph in \cref{fig:conj}.}
    \label{fig:d2o}
  \end{figure}
\end{ex}


\printbibliography

\end{document}